\documentclass[preprint]{vgtc}               

\ifpdf
  \pdfoutput=1\relax                   
  \usepackage{graphicx}                
  \DeclareGraphicsExtensions{.pdf,.png,.jpg,.jpeg} 
\else
  \usepackage{graphicx}                
  \DeclareGraphicsExtensions{.eps}     
\fi%

\graphicspath{{figures/}{pictures/}{images/}{./}} 

\usepackage{microtype}                 
\PassOptionsToPackage{warn}{textcomp}  
\usepackage{textcomp}                  
\usepackage{mathptmx}                  
\usepackage{times}                     
\usepackage{cite}                      
\usepackage{tabu}                      
\usepackage{booktabs}                  

\usepackage{afterpage}
\usepackage{rotating}

\usepackage{xcolor}
\usepackage{enumitem}
\usepackage{fontawesome}
\usepackage{colortbl}
\usepackage{float}

\onlineid{}

\vgtccategory{}

\vgtcinsertpkg

\title{Why More Text is (Often) Better:\\ Themes from Reader Preferences for Integration of Charts and Text}

\author{Chase Stokes\thanks{e-mail: cstokes@ischool.berkeley.edu} %
\and Marti A. Hearst\thanks{e-mail:hearst@berkeley.edu}} %
\affiliation{\scriptsize School of Information \\University of California, Berkeley}

\abstract{
Given a choice between charts with minimal text and those with copious textual annotations, participants in a study \cite{stokes2022} tended to prefer the charts with more text. This paper examines the qualitative responses of the participants' preferences for various stimuli integrating charts and text, including a text-only variant. A thematic analysis of these responses resulted in three main findings. First, readers commented most frequently on the presence or lack of context; they preferred to be informed, even when it sacrificed simplicity. Second, readers  discussed the story-like component of the text-only variant and made little mention of narrative in relation to the chart variants. Finally, readers showed suspicion around possible misleading elements of the chart or text. These themes support findings from previous work on annotations, captions, and alternative text. We raise further questions regarding the combination of text and visual communication.

} 

\keywords{Visualization, text, annotation, preference}

\begin{document}

\maketitle

\section{Introduction}

An influential line of thought in the field of information visualization, spurred in part by the writings of Tufte and the data:ink ratio \cite{tufte1985visual} and additionally by the extended influence of minimalist design \cite{obendorf2009minimalism}, emphasizes the need to simplify the display and eliminate unnecessary marks. This design goal --   to strip information down to its essential essence -- is heavily influential in user interface design more broadly and has helped the field achieve important improvements in usability and interpretability. 

However, this view can be taken too far, so much so that the role of annotations, especially textual annotations, has been under-studied in the design of visualizations \cite{stokesgive}. 
In the research literature, there is often an implicit assumption that charts should be seen and not annotated. Few guidelines exist about how, where, and what kind of text annotations should appear. That said, practitioner books do provide helpful methods, examples, and case studies \cite{knaflic2015storytelling,vidyabridget}. 

To address this gap, we and coauthors recently completed a  study that systematically varied the amount and type of textual annotations shown on univariate line charts. we then compared readers' subjective preferences across these designs \cite{stokes2022}. Contrary to minimalist guidelines, our study showed that significant textual annotation was often preferred over more spare designs, resulting in a guideline that states: ``Rather than aiming for maximally minimalist design, annotate charts with relevant text.''

In this current paper, we look in depth at the written opinions of study participants with the goal of understanding why having more textual annotations on the charts was generally preferred over having less. Using a grounded coding technique on more than 2000 comments, we extract three themes that illustrate the role of textual annotations in visualization. 

The primary and most prevalent theme was an elucidation of the tension between the usefulness of adding information to a chart and the tendency for more information to make the chart look too cluttered. We find a strong preference for more information when it contributes context to the chart without redundancy. The second finding was the presence of narrative structure within text and how a preference for narrative can lead some participants to favor textual forms of expression. The third finding casts new light on the circumstances under which a subset of participants spontaneously brought up potential bias in the information displayed.

\section{Related Work}

A few studies have examined the impact of integrating text with visualizations. Text accompanying visualizations impacts the conclusions a reader makes \cite{kim2021towards, stokes2022} as well as their recollection of the topic \cite{kong2018frames, kong2019trust}. When the title of a visualization was slanted to convey only one possible message within the data, readers were more likely to recall the message or topic from the title, rather than other information in the chart \cite{kong2018frames, kong2019trust}. Additionally, text directs reader takeaways when positioned as a caption \cite{kim2021towards} or as an annotation \cite{stokes2022}. This effect is particularly strong when the text contains external information or provides additional context to the data. 

However, in the context of information visualization, this integration of text and visuals is not always \textit{preferred} by a reader. In assessing participant preferences in conversation with chatbots, Hearst \& Tory found almost half of the readers preferred not to see any visual. When examining similar preferences, Stokes et al. found that readers \textit{most} preferred the integration of text and visuals, ranking the chart with the most annotations highest \cite{stokes2022}. The readers' overall preference for textual or visual communication influenced these rankings, such that those who preferred text as a method of communication also ranked an all-text condition higher than those who preferred visual communication overall.

The content of the text matters as well. In one study, when the text acted to `focus' a chart, the chart performed better in memory tasks and in ratings of aesthetics and clarity \cite{ajani2021declutter}. Text can also accompany a chart as a description for screen readers, in the form of alternative text (alt-text). In the context of alt-text, reader preferences for external information varied \cite{lundgard2021accessible}. Blind and low vision (BLV) readers disliked not being able to interpret the information for themselves, while sighted readers found the storytelling component added by this context to be useful for the interpretation of the chart.

Our work expands the examination of reader preferences, focusing specifically on sighted reader preferences for text annotations on univariate line charts. This also includes an all-text version to compare to visualizations \cite{stokesgive}.

\begin{figure*}
    \centering
    \includegraphics[width=\linewidth]{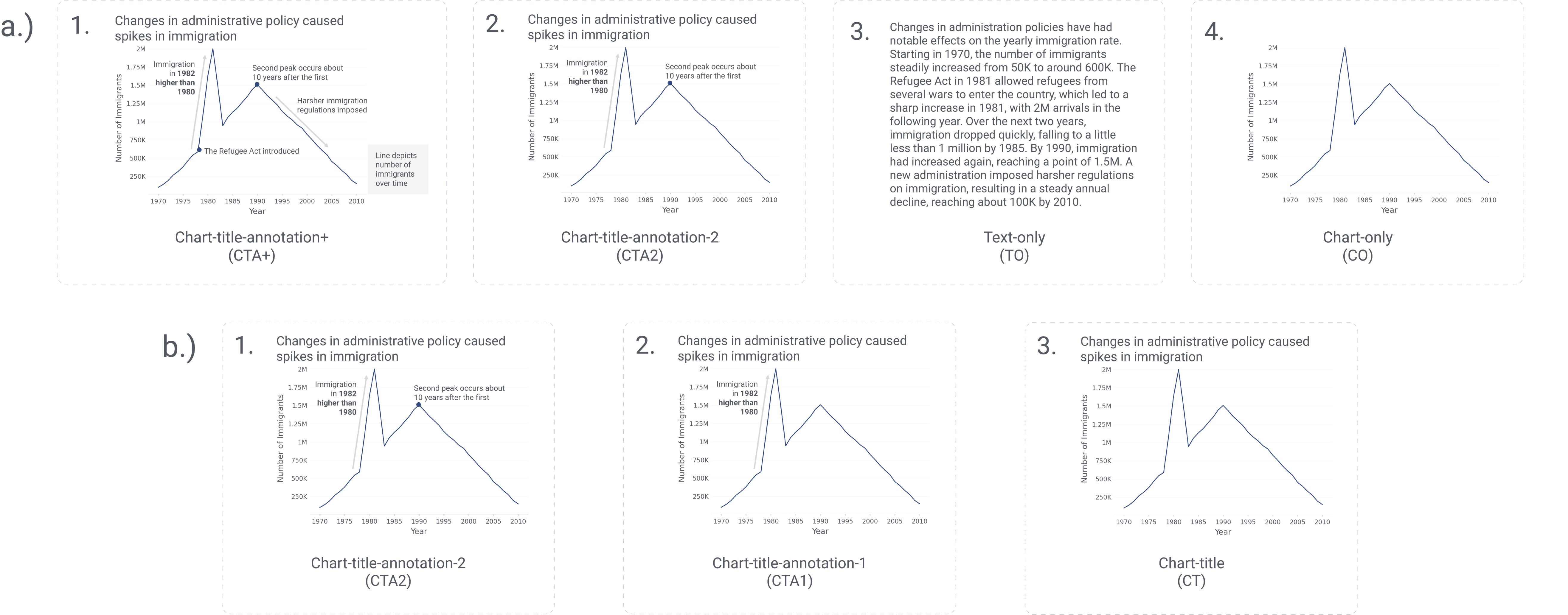}
    \caption{Stimuli sets and corresponding variant labels: (a1) chart-title-annotation+, (a2, b1) chart-title-annotation-2, (a3) text-only, (a4) chart-only, (b2) chart-title-annotation-1, and (b3) chart-title. a) Ranking set A and results \cite{stokes2022}. b) Ranking set B and results \cite{stokes2022}.}
    \label{fig:ranking_sets}
\end{figure*}

\section{Study}


This paper utilizes data that was collected as part of a larger study \cite{stokes2022}, examining the research question, ``What attributes of charts with text do viewers find appealing in comparison to a chart or text on its own? Which do they find unappealing?''

\subsection{Participants}

The original study recruited participants from Amazon Mechanical Turk. Participants were located in the United States, fluent in English, had a 95\% acceptance rate on previous tasks, and completed the survey on a desktop or laptop device. Participants were compensated \$4.00 for a 16-minute survey (\$15 per hour) and were not permitted to take the survey multiple times. 

After filtering participants who did not pass attention checks or who submitted nonsense responses, responses were collected from 302 participants, out of a total 512 recruited. Most participants had completed a 4-year degree and were 35-44 years old on average. On a scale from 1 (overall preference for textual communication) and 6 (overall preference for visual communication) \cite{garcia2016measuring}, participants had an average score of 4.06.

\subsection{Method}

For the full study, participants completed a survey with five main sections. Here, we examine responses from just one of these sections which focused on reader preferences. Further details on the other sections and analyses can be found in Stokes et al \cite{stokes2022}.

\begin{figure}
    \centering
  \includegraphics[width=\linewidth]{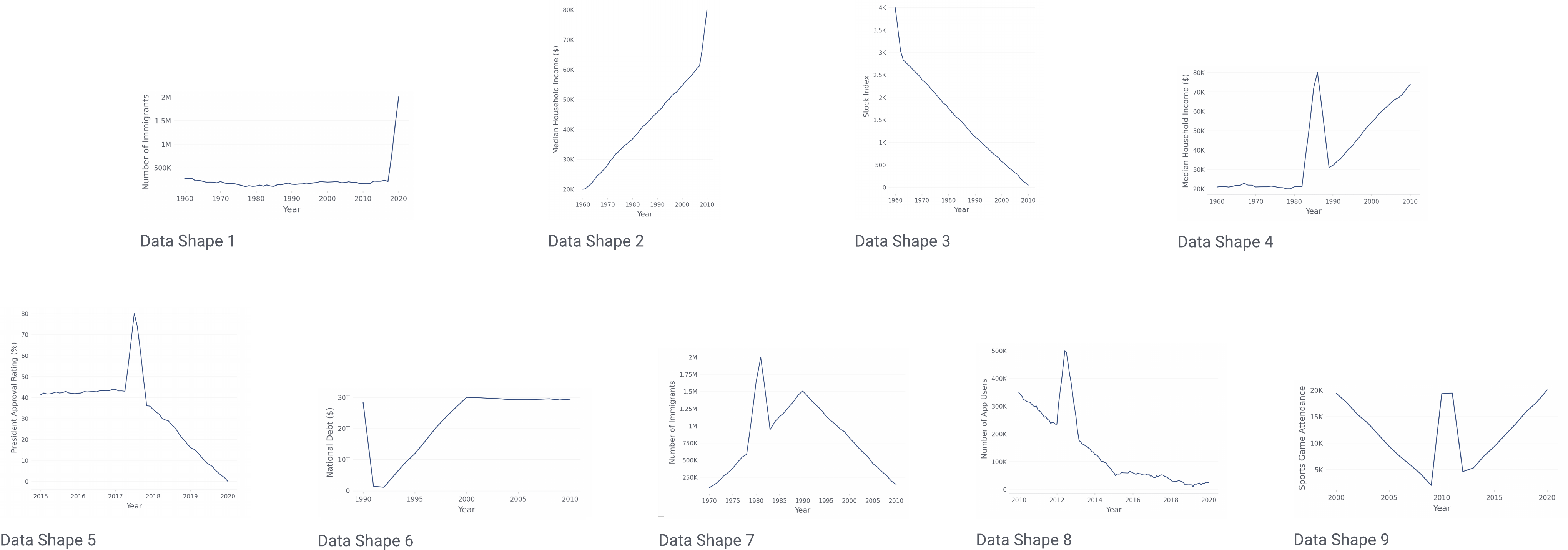}
    \caption{Complete set of possible data shapes for ranking tasks.}
    \label{fig:data_shapes}
\end{figure}

In the survey, participants were shown two different sets of stimuli, labeled A and B (see Figure \ref{fig:ranking_sets}), where A showed variants from 'extreme' ends of the visual to textual range, and (B) showed  a more detailed variation in the number of annotations. There were nine possible data shapes for each ranking set, shown in Figure \ref{fig:data_shapes}. 

A participant first viewed each variant individually and answered a free-response question about which qualities of the information presentation they liked and disliked.  Next, they were asked to rank the set of images ``in the order you would prefer to encounter or see them''.

The stimuli consisted of six total items, shown in Figure \ref{fig:ranking_sets}. Within this paper, we refer to the variants as: chart-only (CO; a chart with only axes labels and tick marks), chart-title (CT; addition of a title), chart-title-annotation-1 (CTA1; addition of an annotation), chart-title-annotation-2 (CTA2; addition of a second annotation), chart-title-annotation+ (CTA+; a chart with a narrative about the data, ranging from 3-6 annotations), and text-only (TO; text passage describing the chart displayed).

\subsubsection{Semantic Content}

The textual annotations used for the stimuli abide by the framework of the four semantic levels from Lundgard \& Satyanarayan \cite{lundgard2021accessible}. Semantic Level 1 (L1) refers to elemental or encoded aspects of a chart. For example, ``\textit{Line depicts number of immigrants over time.}'' Semantic Level 2 (L2) refers to statistical or relational components. For example, ``\textit{Immigration in 1982 higher than in 1990.}'' Semantic Level 3 (L3) refers to perceptual or cognitive aspects. For example, ``\textit{Second peak occurs about 10 years after the first.}'' Semantic Level 4 (L4) provides external context to the chart. For example, ``\textit{Changes in administrative policy caused spikes in immigration.}''

Examples above can be seen in Figure \ref{fig:ranking_sets}. In the original study \cite{stokes2022}, the semantic content of the annotations was varied, such that some of the annotations were simply describing features or attributes in the chart (L1, L2, L3) while others provided external context (L4) (see \cite{lundgard2021accessible} for details). Text annotations were counterbalanced between stimuli, such that all possible combinations of semantic levels were present.

\subsection{Ranking Results}

Participants ranked chart Set A and B independently; the results are displayed in Figure \ref{fig:ranking_sets}. For Set A, CTA+ received the highest ranking, followed by CTA2. TO was ranked third, and CO was ranked fourth. For Set B, CTA2 was ranked first, followed by CTA1, with CT ranked last.

Overall, charts with more annotations were ranked higher than those with fewer. TO was ranked above CO, indicating preference for text over a visual form with no text. The following analysis investigates details motivating these rankings.

\begin{sidewaysfigure}[p]
    \vspace{10cm}
  \makebox[\linewidth]{
        \includegraphics[width=\linewidth]{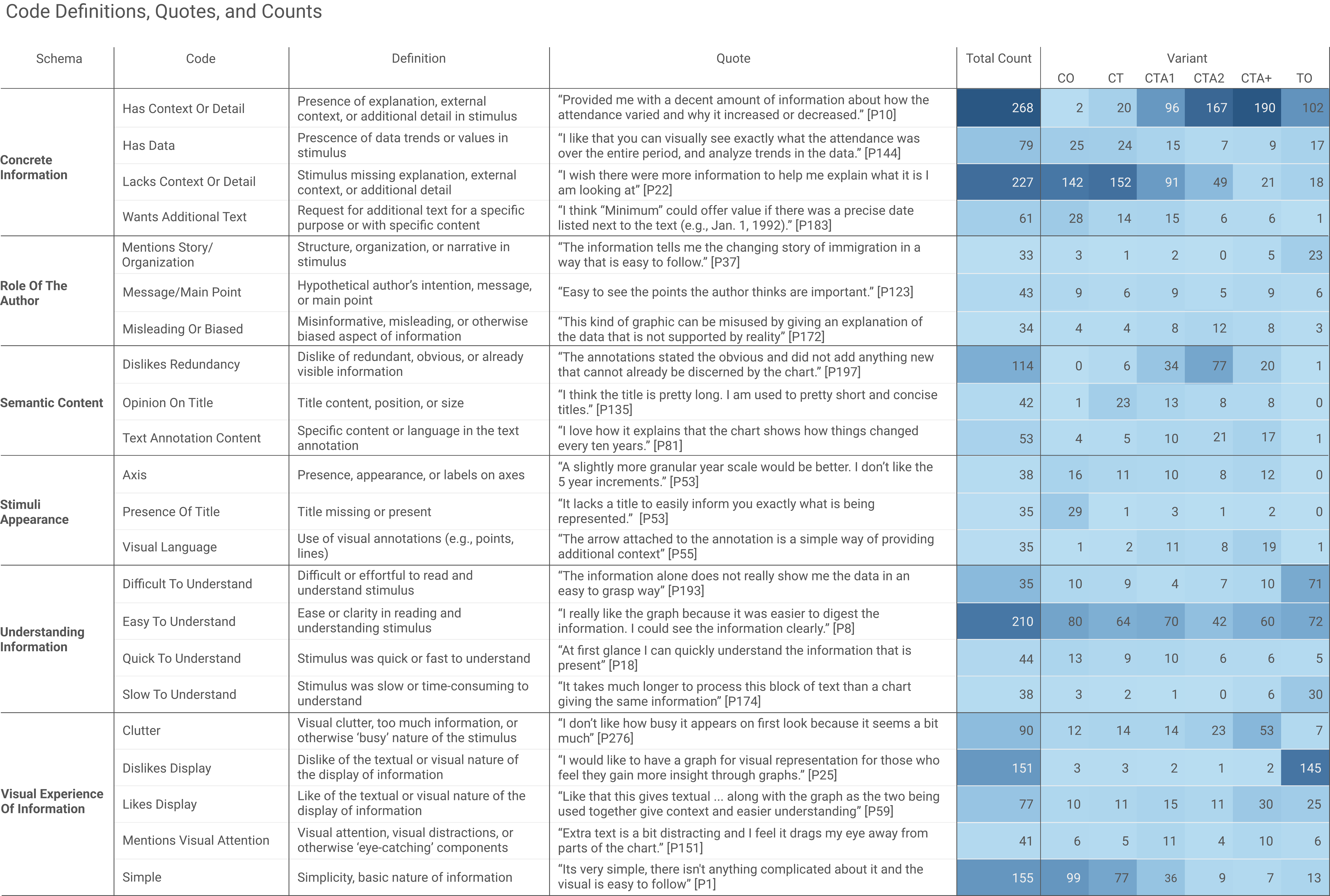}
    }
        \caption{Summary of the thematic analysis. Specific codes are listed, along with their definitions and representative quotes from participants. Schema refers to the broader classifications of codes, which were used to create the themes. The total count of participants who mentioned a code at least once is shown, followed by a breakdown of these counts by stimulus variant.  The darker the shade of blue, the higher the count. Not all participants who mentioned a code did so for all variants; some did so for more than one variant.}
    \label{fig:counts_quotes}
\end{sidewaysfigure}

\afterpage{\clearpage}

\section{Thematic Analysis}

The preference study described above collected a rich set of 2,115 subjective statements about benefits and drawbacks of each chart type.

Using these responses, we conducted a thematic analysis to determine themes arising from participant responses, following the process detailed by Braun \& Clarke \cite{braun2006using, castleberry2018thematic}. This consisted of data familiarization, in which both authors read through responses to form and discuss initial impressions. The first author examined the full dataset and systematically coded the data using the MAXQDA software \cite{maxqda2022}. This process consisted of open coding using an inductive (data-driven) approach, such that the codes and themes identified are linked primarily to the data rather than driven by existing theory \cite{clarke2015thematic}.

The first author then generated initial themes by clustering together similar and related codes. These themes were reviewed by both authors, first in relation to the relevant codes and then in relation to the dataset as a whole. Finally, the themes were more clearly defined and named. These finalized themes are discussed below. Counts for CTA1 are averaged between the two instances (Set A and Set B), rounded up when necessary. All counts are out of the total number of participants (302) unless otherwise stated. Further information on codes and their counts can be seen in Figure \ref{fig:counts_quotes}.

Codes discussed in this paper were mentioned by more than 10\% of participants (31/302) unless stated otherwise. Counts and supporting evidence in the form of participant quotes are given throughout the paper and in Figure \ref{fig:counts_quotes}.

\subsection{Clutter or Context}

\label{section:clutter}

Participants frequently commented on the `simple' (155) or `cluttered' (90) nature of the chart. However, `simple' did not necessarily mean `good,' and `cluttered' did not always mean `bad'.

Participants preferred additional context rather than lacking information in how to approach and analyze the data shown. This was the most frequent comment, with 268 participants appreciating the additional context and 227 disliking charts that lacked context or detail. 114 participants also commented on the presence of `redundant' text, referring to L1, L2, or L3 annotations. This text described what was already visible to the reader in the data, rather than providing new information. While charts with more annotations were seen as more cluttered, the issue that participants cared about most was not the presence of text, but rather the purpose it served. 

\subsubsection{CO and CT}

CO was the variant most frequently referred to as `simple' or `clean' (99), followed by CT (77). P25, in response to CO, wrote ``\textit{I like that the method is almost as simple as it can be.}'' Another, responding to CT, said ``\textit{I like how economical this one is - no annotations.. just the line visual... Simple,} '' [P191].

Both variants were also seen as easy to understand (80; 64). Participants made 49 comments about variants being `quick to understand', with CO receiving 13 of them, the most of any variant. CT was also seen as relatively quick to understand (9/49). One participant noted, ``\textit{[The chart is] pretty straightforward and doesn't take long to understand what it's showing,}'' regarding CT [P3].

Despite these advantages, both variants had serious drawbacks as well. The most frequent comment for both was that the charts lacked context or detail (143; 152). P64 described this absence in CT with ``\textit{I like that I have an image of immigration numbers over time, but I feel like it's not even trying to give me context}.'' 
Out of 70 comments requesting additional text to be added, CO received 29, the most of any variant. CT received 14. Mostly, participants were seeking further information, ``\textit{I don't like that there are no annotations that might explain the spikes and upward/downward trends,}'' [P212].

In the case of CO, participants were also responding to the lack of a title. P92 wrote, ``\textit{The title is missing. It needs a text title to explain the chart.}'' 29 participants commented on this, with only a two participant overlap from the set of 29 which asked for additional text to be added. This represented deeper issue than simply missing further information. It seemed to violate a schema or expectation participants had for a chart: charts have titles. The text in a title served to provide context, but it also served to situate the reader as to what they are looking at.

\subsubsection{CTA1}

CTA1 provided the participant with a title and single text annotation. The chart itself was less `simple,' (36) but the information was still easy to understand (70). P158 responded, ``\textit{I like that there are extra facts and text elements to give context to the data visualization. It helps to understand it better and more easily.}''

CTA1 did provide more context to the information, but for some, that still wasn't enough. CTA1 recieved almost as many comments that it contained context or detail (96) as it did that it lacked context or detail (91). P4 described this tension, ``\textit{The title text combined with the added explanation is good, but I'd have questions about why it continued to decrease like that.}'' Although CTA1 was able to shed some light on the context of the data, it could not provide a full picture with a single annotation.

Additionally, while comments around clutter remain low (14; 12 for CO, 14 for CT), participants were taking note of the semantic content of the text. As described earlier, some of the text annotations simply described the chart, while others provided external context. In comparison to CT (6), CTA1 received more comments (34) disliking redundant information in the annotation. For example, ``\textit{Dislike that the maximum is annotated - that is visually obvious, text adds no information,}'' [P106].

CTA1 provided the most visual balance between clutter and simplicity; comments about clutter increased and comments about simplicity decreased as more text was added to the charts. 

\subsubsection{CTA2}

With a title and two annotations, more participants, though still not many, saw CTA2 as cluttered (23). CTA2 was also the least frequently seen as easy to understand (42). However, roughly the same amount of participants said it was \textit{difficult} to understand (7) in comparison to CT (9) or CTA1 (4). 

P27 indicated a potential issue causing the feeling of `clutter' - more redundant text. They responded, ``\textit{The `Minimum' and other annotated text makes it feel cluttered and overwhelms me}''. 77 participants disliked the presence of redundant text, twice the amount that commented on it for CTA1, the most for any variant.

However, as shown in Figure \ref{fig:ranking_sets}, CTA2 was ranked above CTA1. A large motivation behind this may link to the additional piece of information or detail provided by the second annotation. 167 participants commented on the presence of context or detail, compared to 49 who commented on a lack of such. P5 summed this up well, ``\textit{I like this one even more [than CTA1] because it interprets some of the data and gives a reasoning of the spike}.''

\subsubsection{CTA+}

CTA+ was the highest ranked variant in ranking task A. It provided the most context and detail to the data (190), while still being easy to understand (60). P20 captured this combination, ``\textit{It's an easy to read chart with text explanations at the points of interest. This is definitely preferable because it is both concise yet thorough.}''

Although the annotations performed an important role, 53 participants still commented on the clutter of the chart. The annotations were conveying a great deal of information, but this felt like ``\textit{a lot to take in}'' and ``\textit{very busy}'' [P36]. Rather than stemming from the content of the text, this assessment of clutter likely stemmed from the amount of text. Only 20 participants commented on redundancy in text, the lowest since CT (6). 

Part of this perceived clutter may have also stemmed from the visual language, as 19  participants felt the arrows and points were somewhat redundant. P223 called them ``\textit{a bit pointless and provide unnecessary visual clutter.}'' In particular, the gray arrows were seen as unnecessary, as the direction of the trend was obvious, and they took up more space than the point visuals. 

Despite possible clutter, 30 participants liked the display of the information, ``\textit{I like that this offers both a visual and written explanation of what is going on}'' [P45]. The balance of mode of communication was beneficial, as participants could view the data while also receiving important context.

In summary, the integration of charts and text, when done intentionally and well, has the ability to communicate information in a way which feels understandable despite evoking a sense of clutter.

\subsection{Narrative Organization}

Participants tended to address TO differently than the chart variants. They focused on ways in which text communicates information, and instead of commenting on the visual display, they remarked on the lack thereof. 

\subsubsection{TO}

TO did not receive many comments regarding `clutter' (7) or `simplicity' (13) in display. Participants used different ways to talk about the advantages and disadvantages of textual communication. In particular, the primary drawback of communicating through text was the lack of a visual, coded as `dislikes display method' (145).

Rather than receiving the trends and data points in a chart, participants felt as though they had to visualize the information themselves and hold it in their mind as they read the text passage. P37 described this, ``\textit{It is hard to keep the years and number of immigrants straight in my mind. This is info that would be better served by being displayed visually}.'' This extra ``\textit{mental effort}'' [P41] had evenly split support between participants who found the text passage easy to understand (72) and those who found it difficult (71) or slow (30) to understand. More participants commented on the difficult and slow nature of text communication than any other variant.

However, participants also revealed an advantage of textual communication: the ability to weave a cohesive narrative. Although engaging with the text might be more effortful than engaging with a chart, TO received the most comments about organization and structure (23/33). 21 of these 23 participants \textit{liked} the organization; ``\textit{this narrative clearly describes change over time... including reasons for the causation}'' [P134]. The temporal order of the story text made sense, even if visualizing the precise quantities was taxing. Despite the effort to convey the same information, there were few comments about any kind of narrative or organization for CTA+ (5).

While most participants preferred a form of multi-modal communication, 25 stated they preferred the text display: ``\textit{I like the nuance and detail allowed by language},'' [P169]. These responses illustrate the unique affordances of text communication. Simply conveying the same information in a chart does not necessarily convey the narrative component. None of the variants contained a particular focus on data-storytelling, but the text passage was not written to explicitly tell a story either. Rather, it was meant to be a text variant expressing the data and context of the chart. Text communication of data and context provided a story or narrative to the data more inherently than a chart with the same information.

\subsection{Misleading and Manipulating}

Finally, some participants felt cautious or suspicious of the stimuli presented, with 34 respondents mentioning a misleading or biased component of the variants. These comments were mostly found regarding charts with annotation (28/39 comments) and less around more single-modal communication (11/39).

\subsubsection{TO}

As mentioned previously, participants didn't comment on TO in the same way they did for chart variants. Very few mentioned anything about bias - the presence (3) or the lack (3).  While participants seemed aware of the hypothetical author's role in constructing the narrative around the text, they were not as likely to comment about the possible intentions or the ability to mislead a viewer with text.

\subsubsection{CO and CT}

Of the chart variants, CO (4) and CT (4) received the least comments regarding possible misleading or biased information. Only 18 participants commented on the lack of opinion or bias present in stimuli, meaning it did not reach the 31 participant minimum to be considered a significant code. However, in this context, it provides a useful perspective into why these variants may have received lower rates of `misleading' comments.

In total, 26 comments were made about the \textit{lack} of bias present. Of these, 11/26 referred to CO and 7/26 referred to CT, more than any other variant. The issue posed by the lack of context became a positive attribute in this area, as participants felt, ``it just presents the data,'' [P65] and they could ``make [their] own determination as to what is happening,'' [P113]. The absence of context or additional information meant that participants could select for themselves what they believed to be the important or meaningful areas of the chart.

Of the few (4) comments about bias in CO, they mostly concerned the artificial, ``bugged'' appearance of the data [P246]. This was a component of the stimuli, since they were artificially generated to have strong features. In the case of CT, participants felt the title text was misleading or ``injecting an opinion/political view,'' [P187]. Many of the comments of bias arose around text, making it more of an issue in charts with more annotations.

\subsubsection{CTA1, CTA2, and CTA+}

Comments about bias or misleading information were more common in the charts with more annotations: CTA1 (8), CTA2 (12), and CTA+ (8). Issues with CTA1 stemmed primarily from the lack of additional context provided. By only providing a single annotation, the chart seemed to take on a narrow focus, with one participant commenting ``\textit{hard to believe that that should be the key takeaway given the rest of the chart},'' [P123].

With an additional annotation, in CTA2, some participants still found issue with the annotations provided, ``purposely trying to lead me into a direction and viewpoint without giving me enough details,'' [P65]. While the charts provided context, participants still felt as though they were not seeing the full story - biased as much for the context shown as for the context omitted. 

CTA+ included more context to the chart and received only one comment that there may be important information omitted [P6]. The other 7 participants found the text provided too much interpretation of the data, as  ``\textit{It makes conclusions based on opinion. It thinks for the reader instead of allowing the reader to do the thinking,}'' [P54]. This supports the sentiment found in the initial examination of this level taxonomy \cite{lundgard2021accessible}, as BLV readers disliked the level of interpretation L4 text provided.

\section{Explicit Comparisons}

In addition to commenting on what they liked and disliked about each variant, participants were also provided a text box to elaborate on their reasoning for the ranking of each stimuli set shown in Figure \ref{fig:ranking_sets}. This response was optional; 200 participants responded to one or both opportunities for elaboration.

The first author coded these responses with the same coding scheme used for the like/dislike responses. Any counts given are out of the 200 participants who responded, unless otherwise specified.

Participants most frequently used the existence of context or detail to rank the stimuli, with 159 mentioning that they ranked a variant higher because it provided information or context and 70 mentioning that they ranked a variant lower because it lacked the context provided by another. Another key component of the ranking choices was the variant being `easy' (75) or `quick' (23) to understand, which sometimes (45) overlapped with the presence of context. For instance, P8 wrote, ``\textit{When the graphs have some text explaining the rise or fall, it gives more context. It makes it easier to understand}.''

The visual or textual nature of the variant also played a role in the ranking, as 67 participants mentioned ranking a variant higher because they preferred the way it was shown over another possible medium. These comments were more common when ranking Set A (61) than when ranking Set B (11), as Set A contained the TO variant which led to comparisons between methods of communication.

Ranking Set B, on the other hand, led to more fine-grained comparisons, particularly focusing on the content of the annotations. While 51 participants commented on their dislike of the redundancy of annotations as a primary reason for ranking a variant lower, Set B received far more of these comments (43) than Set A (13). The primary differences between variants in Set B were the presence of an additional one or two annotations, which meant that most of the participant reasoning centered around the impact of each annotation. At times, the annotation added context, but it also may have added redundant information.

Participants made relatively few comments regarding simplicity (33) and clutter (27) in comparison to the responses describing likes and dislikes. While these factors may have come up in how they felt about individual variants, they were not significant enough to impact the ranking of the sets as a whole. This further supports Section \ref{section:clutter}, as one participant put it, ``\textit{While [Rank] 1 is too busy, it is also the most informative},'' [P55]. The clutter of information was a minor issue compared to the benefit it provided.

\section{Discussion}

By utilizing multiple methods of communication, combining text and charts often allows for a greater amount of information to be conveyed. Further understanding participant preferences for this combination allows designers and researchers to better assess the impact of this additional information.

Common visualization design guidelines promote practices of minimalism, which avoids overwhelming the reader with `clutter.' The responses examined in this paper indicate that the presence of `clutter' does not necessarily make for a poorly designed visualization. Instead, readers were able to appreciate the \textit{purpose} served by text annotations and preferred to receive this information and additional context. Readers generally showed a strong dislike of redundant information which did not serve to further their understanding of the data.

`Clutter' was generally a disadvantage of the visualization, but the context and detail the annotations provided made a key difference between otherwise identical charts. This supports Guideline 1 from Stokes et al., ``Rather than aiming for maximally minimalist design, annotate charts with relevant text.'' \cite{stokes2022}. Trading simplicity for information is advantageous for reader preferences.

Text communication is less subject to the issue of `clutter' but can be more difficult or slower to understand. The lack of visual means more effort on the reader's part to visualize information on their own, but text can provide a story or narrative structure which may be more difficult to convey in a single, static visualization. In this examination of reader preferences, including a text-only variant of the information allowed for a clearer understanding of the importance of visualizing the information and the drawbacks of using a single method of communication.

The authors of this paper found the creation of the text-only variant to be a useful exercise in creating and structuring a narrative around the data. This narrative, which was explicitly noticed by a subset (7.6\%) of participants provided a useful starting point for the creation of annotations. This also provides support for Guideline 4 from Stokes et al., ``Consider a text-only variant that can stand alone.'' \cite{stokes2022}. Text offers a different set of advantages and disadvantages than visual, which are important to consider when combining the two.

Text was also less considered a candidate for bias. When the same information is shown as an annotation on a chart, comments about bias or misleading text are more common. Visualizations can be structured in misleading ways \cite{cairo2019charts} or be used as a vector for misinformation \cite{lee2021viral}. A minority (11.3\%) of readers seem aware of this possibility and proceed with caution around context shown or not shown in a visualization. However, the text was presented in a neutral way; if it had been deliberately written in an opinionated manner, responses about bias in text may have been generated.

These results emphasize the importance of understanding the role of `trust' in relation to visualized data, especially as researchers turn to visualization to convey information regarding machine learning \cite{chatzimparmpas2020state, xiong2019examining}. It remains unclear which components of a visualization (e.g., amount of information, transparency, etc.) motivate trust.

Reader preferences in this paper illuminate how a reader engages with data or information and how they view the hypothetical author of the visualization. Informing the reader should be a top priority as it received the most comments. Narrative visualizations and trust in visualizations are two areas of important further investigation, as indicated by the themes extracted in this paper.

\section{Limitations}

Themes from this paper only refer to preferences, which may not correspond to performance. The annotations which are preferred by participants do not necessarily lead to a better understanding of the data, despite readers thinking that the visualization is easier to understand. Further work would be necessary to evaluate reader performance and understanding with a visualization.

These preferences also only refer to temporal data plotted in a line chart. Other chart types and other types of data are not yet explored. Future work regarding preferences for text and chart integration should examine additional types of data and visualization to assess additional factors not present in the cases studied here.

The charts in this study were presented in a stand-alone manner. It could be that if they had been embedded into documents and surrounded by paragraphs explaining the charts, the preferences might be different. 

Finally, this analysis can only examine what participants \textit{said}, not what they \textit{did not say}. For example, just because a reader did not find a variant to be misleading does not mean they found it to be trustworthy. This would require more in-depth questioning on particular topics to account for in future work.

\section{Conclusion}

Reader opinions explored in this paper illuminate key contributions of textual annotations in visualization. While more annotations may lead to more clutter on a chart, they also lead to more information expressed. This additional information was often useful to the reader and outweighed the possible harm of clutter when contributing context. Conveying the same information in text received a greater amount of comments regarding narrative and structure, further demonstrating the possible benefits of text communication compared to visualizations. Text annotations may signify greater risk for bias expressed within the chart, as chart variants with more annotations received more comments regarding misleading information. These findings encourage continued investigation of the integration of text and charts.

\acknowledgments{
This research was supported in part by a gift from the Allen Institute for AI.
}

\bibliographystyle{abbrv}

\newpage

\bibliography{bib}

\end{document}